\documentclass{egpubl}
\usepackage{eurovis2017}
\ConferencePaper        
\SpecialIssuePaper         
 \electronicVersion 

\ifpdf \usepackage[pdftex]{graphicx} \pdfcompresslevel=9
\else \usepackage[dvips]{graphicx} \fi

\PrintedOrElectronic

\usepackage{t1enc,dfadobe}

\usepackage{egweblnk}
\usepackage{cite}
\usepackage{color}
\usepackage{xspace}
\usepackage{enumitem}
\usepackage{amsmath}
\usepackage[final]{review}

\def \prototypename {Graffinity\xspace}
\def \connectivitymatrix {connectivity matrix\xspace}
\def \intermediatenodes {intermediate node table\xspace}

\title[Graffinity: Visualizing Connectivity in Large Graphs]%
      {Graffinity: Visualizing Connectivity in Large Graphs}

\author[Kerzner et al.]
       {
        E. Kerzner$^{1}$ A. Lex$^{1}$ C.L. Sigulinsky$^{1}$ T. Urness$^{2}$\and
         B.W. Jones$^{1}$ R.E. Marc$^{1}$ M. Meyer$^{1}$\\
         $^1$University of Utah, USA\\
         $^2$Drake University, USA
         \vspace{-10mm} 
       }

\begin{document}

\teaser{
 \includegraphics[width=\linewidth]{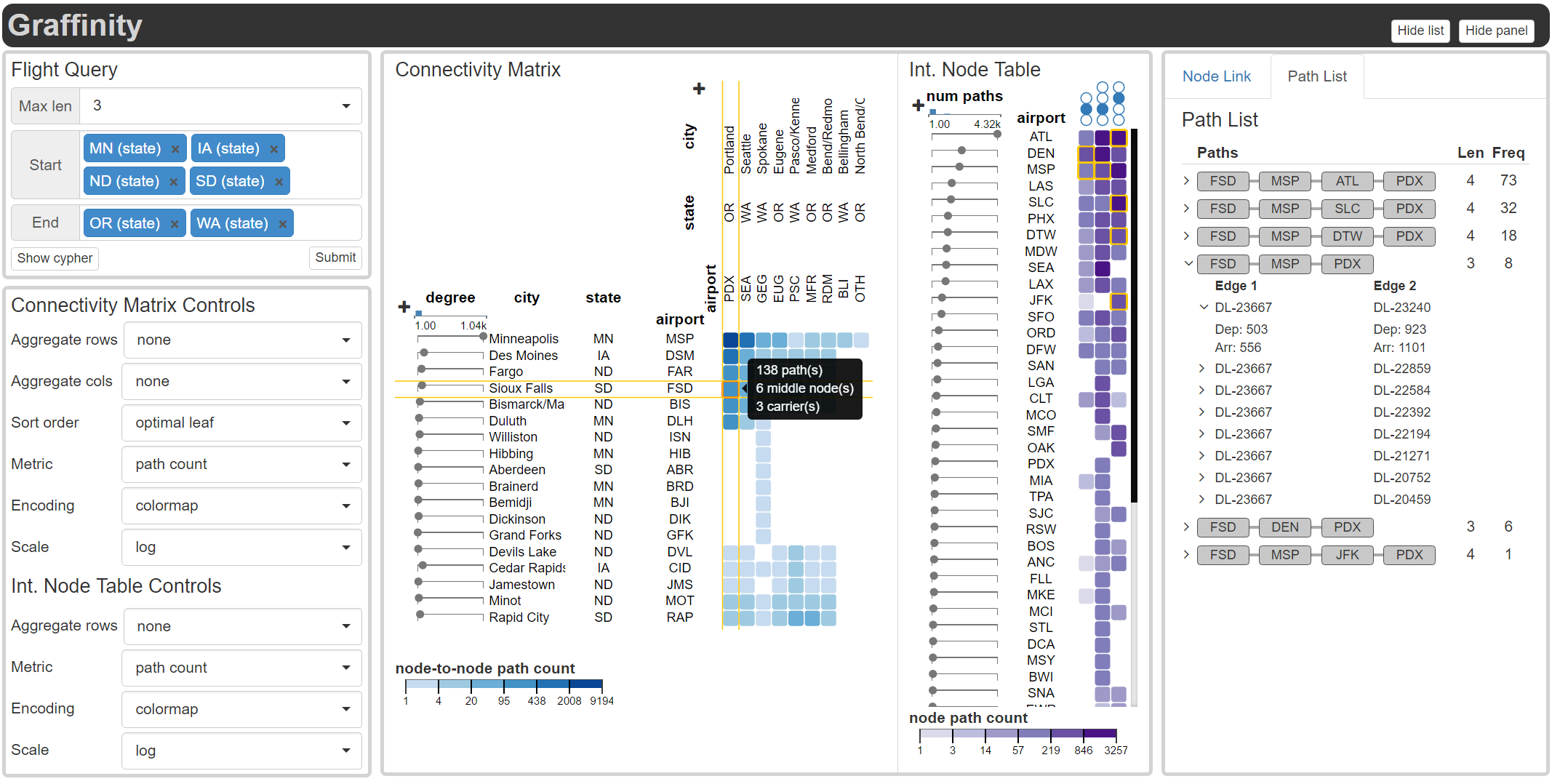}
 \centering
 \vspace{-5mm}
  \caption{\prototypename visualizing 11727 flight paths with length $\leq 3$ connecting states in the mid-western USA (Minnesota, Iowa, North Dakota and South Dakota) to states in the Pacific Northwest (Oregon and Washington). \prototypename consists of five views: the query interface, the \connectivitymatrix, the \intermediatenodes, and two views showing details about selected paths: the path list and the node-link view.  The 138 paths connecting the airport FSD (Sioux Falls, SD) to PDX (Portland, OR) are selected and displayed in the path list view.}
\label{fig:full-tool}
}

\maketitle

\begin{abstract}
Multivariate graphs are prolific across many fields, including transportation and neuroscience. A key task in graph analysis is the exploration of connectivity, to, for example, analyze how signals flow through neurons, or to explore how well different cities are connected by flights. While standard node-link diagrams are helpful in judging connectivity, they do not scale to large networks. Adjacency matrices also do not scale to large networks and are only suitable to judge connectivity of adjacent nodes. A key approach to realize scalable graph visualization are queries: instead of displaying the whole network, only a relevant subset is shown. Query-based techniques for analyzing connectivity in graphs, however, can also easily suffer from cluttering if the query result is big enough. To remedy this, we introduce techniques that provide an overview of the connectivity and reveal details on demand. We have two main contributions: (1) two novel visualization techniques that work in concert for summarizing graph connectivity; and (2) Graffinity, an open-source implementation of these visualizations supplemented by detail views to enable a complete analysis workflow. Graffinity was designed in a close collaboration with neuroscientists and is optimized for connectomics data analysis, yet the technique is applicable across domains. We validate the connectivity overview and our open-source tool with illustrative examples using flight and connectomics data.

\begin{classification} 
\CCScat{Computer Graphics}{H.5.2}{Information Interfaces and Presentation}{User Interfaces}
\end{classification}

\end{abstract}

\section{Introduction}

Graphs are an important datatype across many domains from transportation to neuroscience. Graph nodes represent entities and edges the connections or relationships between those entities. For instance, graphs can model the flights (edges) between airports (nodes) or synapses (edges) between neurons (nodes). In multivariate graphs, both nodes and edges can be associated with categorical attributes, such as the city of an airport, and quantitative attributes, such as the size of a synapse. Analyzing multivariate graphs often involves understanding some combination of the graph's topology and attributes~\cite{Lee2006,VandenElzen2014}.

One important area of graph analysis is concerned with examining the direct and indirect connections between entities, their connectivity. This is important for understanding structures implied by the graph's topology, such as airline routes that directly or indirectly connect cities. Understanding the direct and indirect connections involves analyzing a combination of the graph's adjacency (direct connections), connectivity (presence of paths connecting entities), and accessibility (entities reachable from a certain one)~\cite{Lee2006,Pretorious2013}. In this paper, we use the term \emph{connectivity} to refer to the direct and indirect connections between entities based on paths, potentially considering node and edge attributes.

Understanding the connectivity of a graph is challenging because the number of possible paths connecting two entities increases exponentially with graph size~\cite{Barabasi2016}.  This scalability problem is exacerbated by standard graph visualizations such as node-link diagrams and adjacency matrices which have their own limitations when used for connectivity analysis. Node-link diagrams excel at topology-based tasks for small graphs but degenerate to hairballs for larger graphs~\cite{Shneiderman2006a}. Adjacency matrices are slightly more scalable for tasks related to adjacency in large graphs, but are ill-suited for tasks involving indirect connectivity because they require tracing across rows and columns to follow paths~\cite{Ghoniem2005}. As the size of a graph increases, specialized techniques are needed to make sense of its connectivity.

Query-based approaches (e.g., \cite{Aleman-Meza2005, Heim2009, Zhang2013, Tu2013, Partl2016}) are helpful when dealing with large graphs in general, and for understanding graph connectivity in particular. These systems allow analysts to query the graph for the connections between a set of nodes and return a subset of the entire graph. These subsets are often displayed as lists~\cite{Partl2016}, subgraphs~\cite{Heim2009}, or use a specialized representation~\cite{Tu2013}. But as the size of query results increases, analyzing connectivity again becomes challenging due to the large number of potential paths.

In this paper we propose a new technique for making sense of connectivity in large graphs. Our technique provides a flexible overview of path-based connectivity, enabling a user to explore interesting subsets of paths in a highly scalable way. The design of the technique was motivated by a collaboration with neuroscientists which, along with a review of visualization literature, allowed us to identify a set of design requirements for summarizing graph connectivity in a query-based system.

Based on these requirements we present two contributions: (1) two novel and complementary visualization techniques for summarizing the connectivity in a subset of a graph selected by queries, the \connectivitymatrix and the \intermediatenodes; and (2) \prototypename, an open-source implementation of these techniques. Although this work is motivated by our collaboration with neuroscientists, our visualization techniques and prototype generalize to graph analysis in other domains. We validate this work through illustrative examples and case studies with flight and neuroscience data.
\section{Requirements}
We introduce a set of requirements (\ref{req:query}-\ref{req:context}) for visualizations designed to summarize graph connectivity. We identified these requirements in a user-centered design process involving a group of up to eight neuroscientists over a period of 18 months. We used methods including contextual inquiry~\cite{Holtzblatt1993}, creativity workshops~\cite{Goodwin2013}, and informal interviews to elicit requirements and receive feedback on prototypes. The requirements were also influenced by prior visualization research, discussed in Section~\ref{sec:related}. 

While these requirements were informed by a domain collaboration, we argue that they apply broadly. They are, however, not meant to be exhaustive for general graph analysis, but are targeted at a use case of analyzing connectivity between node sets. This so-called many-to-many analysis is useful for understanding relationships in a graph at a higher level of abstraction than individual nodes. For instance, an airline analyst may be interested in how two states, both with many airports, are connected by air travel. Another example is the analysis of trade or migration between geographic regions that are represented as sets of nodes~\cite{Yang2016}. In neuroscience, researchers examine the flow of signals between different types of neurons~\cite{Lauritzen2016}. 

Many-to-many analysis is often performed on graphs that are too large to be drawn directly. In these cases, analysts often use queries to identify interesting subgraphs~\cite{VandenElzen2014}. Hence, our requirements focus on query-based connectivity analysis between node sets. 

We assume that all measures of connectivity are based on short paths connecting the nodes. \newrem{It follows that}{Hence} our requirements address both abstract measures of connectivity as well as specific paths connecting the nodes.

\begin{enumerate}[label=\textbf{R\arabic*}]
\setlength\itemsep{0.5em}

\item \label{req:query} \textbf{Query many-to-many paths.} Analysts should be able to specify path queries based on node lists, shared attributes of starting or ending nodes, or the types of nodes and edges involved in the paths.

\item \label{req:overview} \textbf{Visualize an overview of connectivity.} Analysts should be presented with a visual summary of the relationships between the nodes that they queried for. It is important that this representation appropriately scales to handle large numbers of paths. Connectivity can also be defined in various ways, hence a system targeted at analyzing connectivity should allow analysts to specify different metrics to represent connectivity.

\item \label{req:aggregate} \textbf{Support dynamic aggregation of nodes and paths.} In order to understand higher-level structures in a network, analysts may be interested in relationships between node sets. To support this type of analysis, dynamic aggregation of nodes, and consequently of the paths connecting these nodes, should be supported.

\vspace{4mm} 

\item \label{req:details} \textbf{Visualize path details.} The details of paths, including the individual nodes and edges that make up paths, as well as the node and edge attributes should be accessible on demand.

\item \label{req:context} \textbf{Visualize path context.} The context of a path describes how it is embedded within the topology of the graph. Also, when appropriate, a meaningful spatial representation of the nodes and edges should be available.

\end{enumerate}

Finally, underlying our requirements is the assumption that analysts have already identified interesting queries about the connectivity. These queries may be based on existing domain knowledge and bottom-up analysis such as tracing paths in node-link diagrams or other visualization techniques discussed in the next section.

\begin{figure}
\includegraphics[width=\linewidth]{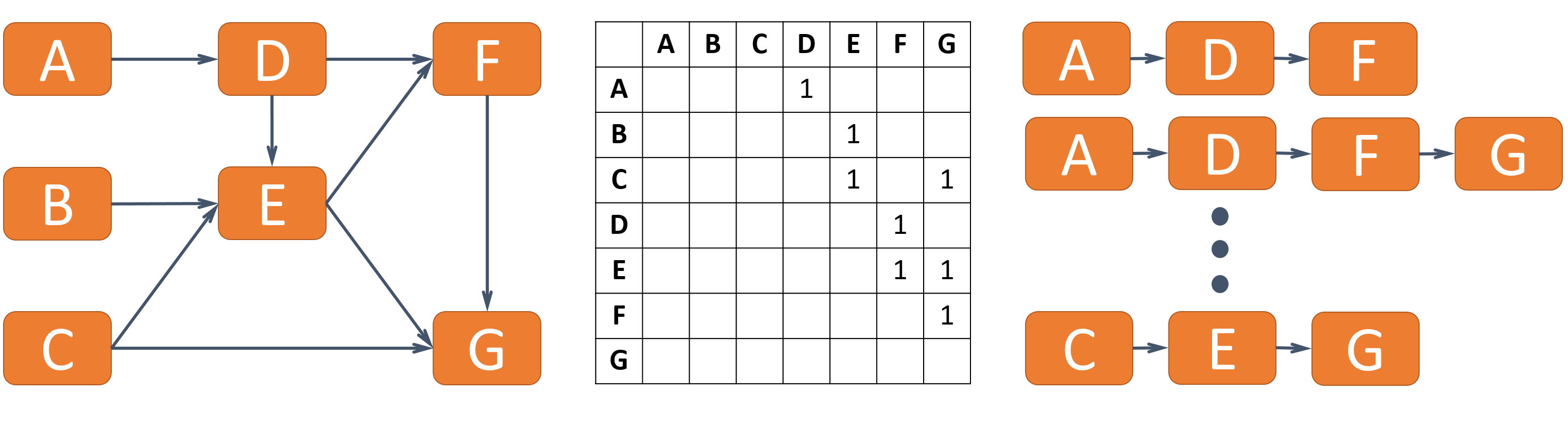}
\caption{Analyzing node connectivity is challenging with traditional graph encodings and path listing techniques. A query for paths connecting nodes A,B,C with nodes F,G returned a subgraph shown here. Node-link diagrams (left) give an overview of graph topology but require manual tracing to analyze relationships between the start and end nodes. Adjacency matrices (middle) are ill suited for connectivity  analysis as tracing paths is challenging in matrices. Path lists (right) do not provide a connectivity overview.}
\label{fig:traditional}
\end{figure}
\section{Related Work}
\label{sec:related}

We focus our discussion of related work on techniques that support path-based connectivity analysis in large, multivariate graphs.  Summaries of the extensive research on graph visualization beyond path analysis are available for various areas, including visualization of large graphs~\cite{VonLandesberger2012}, dynamic graphs~\cite{Beck2016}, and multivariate graphs~\cite{Kerren2014}.

Representations for paths in graphs include traditional node-link layouts, adjacency matrices, and path-listing techniques~\cite{Partl2016} (Figure~\ref{fig:traditional}). Each of these techniques can be combined with an initial query step to reduce a larger graph into a smaller subgraph (\ref{req:query}) to enhance scalability. Path queries are supported by general purpose graph software packages, such as Tulip~\cite{Mutzel2004} and Gephi~\cite{Bastian2009}, and databases such as Neo4j~\cite{Neo2016}.

Traditional node-link diagrams support the exploration of connectivity by enabling analysts to trace paths to identify the relationships between nodes (\ref{req:details}) within the graph's topological context (\ref{req:context})~\cite{Ghoniem2005}. They fail to scale to many nodes and paths (\ref{req:overview}), however, as they require manual tracing of paths, and they quickly degenerate to hairballs when they exceed about 50 nodes and 200 links~\cite{Shneiderman2006a}. RelFinder~\cite{Heim2009} and the path topology view in Pathfinder~\cite{Partl2016} are examples of node-link diagrams being used to display the results of path queries.

Adjacency matrices are generally considered ill-suited for path-related tasks because they require tedious manual tracing between rows and columns to follow the paths~\cite{Ghoniem2005}. Augmented matrices exist to support browsing paths and accessing details of those paths (\ref{req:details}). In MatLink, Henry et al.~\cite{Henry2007} augmented adjacency matrices with additional edge representations. This approach has been expanded by Shen and Ma~\cite{Shen2007} who draw links directly on top of matrices. Recently, the Ego-Lines tool~\cite{Zhao2016} has used a similar approach for representing paths in ego-centric adjacency matrices. These augmented matrix approaches are appropriate for following a relatively small number of paths, but do not provide a scalable overview of connectivity (\ref{req:overview}). 

Matrices can also be augmented to display aggregations of relationships (\ref{req:aggregate}). Aggregated matrices such as those used in Honeycomb~\cite{vanHam2009} and MapTrix~\cite{Yang2016} are suitable for analyzing adjacency between sets of nodes. Yet, these approaches suffer from the same problem as non-aggregated matrices when considering connectivity and hence do not provide an adequate overview of connectivity (\ref{req:overview}).

There are also node-link based aggregation approaches for graphs. PivotGraphs~\cite{Wattenberg2006} create aggregate representations of graphs based on node attributes (\ref{req:aggregate}), but this representation hides the paths that connect individual nodes and hence does not meet the requirements related to individual paths (\ref{req:details}, \ref{req:context}).  GraphCharter~\cite{Tu2013} is a pivot graph implementation modified to support iterative query-based path browsing, but does not adequately provide information about connections between many-to-many nodes. Details-to-overview-via-selection-and-aggregation~\cite{VandenElzen2014} enables users to transform node-link representations of a graph into aggregated summaries (\ref{req:aggregate}). These summaries can explain connections between sets of nodes, but do not necessarily support analysis of connections within those sets, or between individual nodes (\ref{req:details}). 

Statistical summaries can be used to give an overview of connectivity. B-Matrices~\cite{Bagrow2007,Czech2011} and graph prisms~\cite{Kairam2012} offer such summaries of nodes and edges in a graph, such as the number of reachable nodes, but these approaches contain little information about relationships between specific nodes (\ref{req:overview}).

Specialized techniques are particularly suitable for query-based path analysis and have focused extensively on querying paths between a small number of start and end nodes. RelClus~\cite{Zhang2013} clusters paths hierarchically according to length and co-occurring nodes (\ref{req:aggregate}) and displays these clusters in a tree view. This technique, however, does not provide an overview of many-to-many relationships without manual aggregation (\ref{req:overview}). Aleman-Meza et al.~\cite{Aleman-Meza2005} support ranking and browsing paths  (\ref{req:details}) to identify interesting regions of a graph, but do not provide explicit summaries of the resulting connectivity (\ref{req:overview}). PathFinder~\cite{Partl2016} supports querying for paths between sets of nodes (\ref{req:query}) and interactive browsing and ranking of those paths (\ref{req:details}), but does not provide an adequate overview of the connectivity~(\ref{req:overview}).

This work aims to support queries between large sets of start and end nodes, visualize an overview of their connectivity, and then support analysis of the paths in detail.
\begin{figure}
\centering
\includegraphics[width=\linewidth]{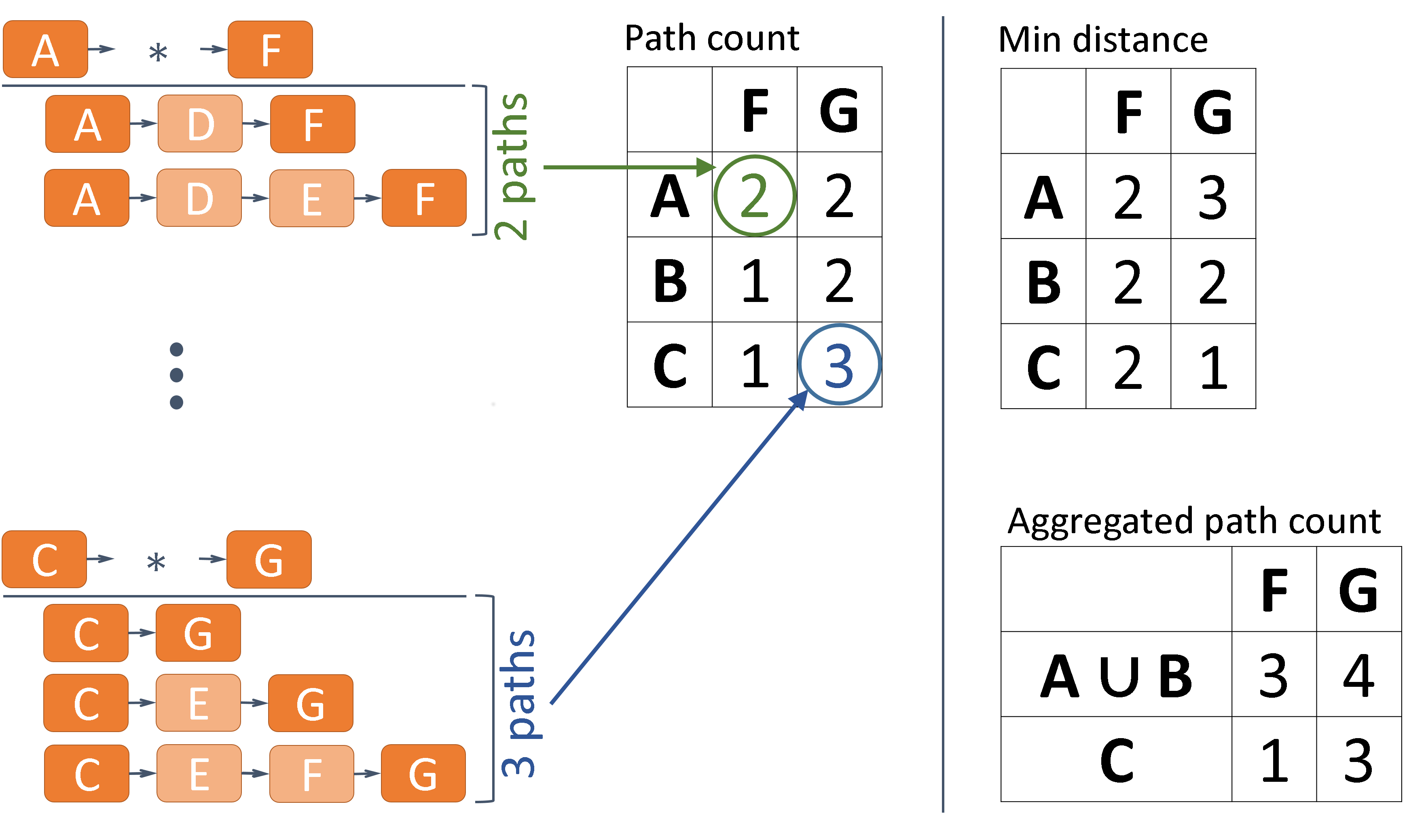}
\caption{The construction of a \connectivitymatrix using the subgraph introduced in Figure~\ref{fig:traditional}. We create path sets based on common start and end nodes then represent those sets in a matrix where each cell shows a metric applied to paths connecting a pair of nodes. Examples of path-based metrics shown here are the count of paths connecting two nodes and the minimum distance between two nodes. Additionally, the matrix rows and columns can be aggregated by computing the union of the corresponding path sets.}
\label{fig:matrix-diagram}
\end{figure}

\section{Connectivity Overviews}
\label{sec:connectivity-overviews}

In this section, we describe two visualization techniques for providing an overview of graph connectivity. These two complementary techniques are designed to give an overview of paths between nodes (\ref{req:overview}) and support dynamic aggregation of those paths (\ref{req:aggregate}). The first technique is the \connectivitymatrix, which provides an overview of paths as relationships between start and end nodes. The second, complementary, technique is the \intermediatenodes, which provides additional details about the role of intermediate nodes in these paths. These two techniques are implemented in a prototype, called \prototypename, that addresses the other requirements of querying for paths (\ref{req:query}), accessing path details (\ref{req:details}), and providing context (\ref{req:context}). We discuss these features of \prototypename in Section~\ref{sec:prototype}. Here we describe the \connectivitymatrix and the \intermediatenodes assuming that a user has provided a query for paths connecting sets of nodes. 

\subsection{Connectivity Matrix}

We designed the \connectivitymatrix to provide users with an overview of path-based connectivity when they query for paths between sets of nodes. The \connectivitymatrix visualizes sets of paths connecting start and end nodes. We apply metrics to these path sets, such as the count of paths, and display the results of these metrics in a matrix. The matrix rows correspond to the start nodes and the columns correspond to the end nodes. This matrix representation is a generalization of the adjacency matrix for showing path relationships. In the remainder of this subsection we provide a definition of path sets, example metrics to analyze those sets\rem{, discuss visual encodings for the metric results}, and discuss the aggregation of paths.

A query returns a subgraph $G=(N,E)$ that contains paths between the user-specified start nodes, $N_{start} = \{start_{0},~start_{1},~\dots\}$ and end nodes, $N_{end} =\{end_{0},~end_{1},~\dots\}$. The paths are $P=\{p_{0},~p_{1},~\dots~,p_{k}\}$. We define \emph{connectivity sets}, $C$, for all pairs of the start and end nodes as the set of paths that connect those nodes. 

Formally,
\begin{equation*}
\begin{split}
C(start, end) = \{p~|~p\in~P~\wedge~Start(p)=start~\wedge~End(p)=end\}\\
~\forall~start \in N_{start},\forall~end\in N_{end} 
\end{split}
\end{equation*}

Each of these sets contains the paths matching the query criteria that connect a pair of start and end nodes. An example derivation of the path sets is shown in Figure~\ref{fig:matrix-diagram}.

Each row in the \connectivitymatrix corresponds to a start node, $start \in N_{start}$, and each column corresponds to an end node, $end \in N_{end}$. Each matrix cell represents the path set, $C(start, end)$. An alternative derivation of this matrix that relies on properties of the graph's adjacency matrix is described in the supplemental material.

We use the cells of the matrix to visualize a metric derived from its path sets. For example, in Figure~\ref{fig:matrix-diagram} (left) we use a metric that counts the number of paths in a set. More generally, a metric is a function that operates on a path set and returns one or more values representing those paths. Two domain-agnostic metrics are the count and minimum length of paths in a set (Figure~\ref{fig:matrix-diagram} (right)), yet there are many possibilities for other metrics that account for node and edge attributes, e.g., taking edge weights into account. 

The result of the metrics can be displayed using various visual encodings. Color coding the cells (i.e., creating a heat map) provides a visual summary of connectivity when using metrics that return a single value per set. More complex metrics that return an array of values could make use of a small multiples display of the table or a glyph representing multiple values in a cell~\cite{Elmqvist2008}. These are described in more detail in Section~\ref{sec:prototype}. 

\rem{As the interpretation of matrices depends strongly on the order of its rows and columns, the \connectivitymatrix should also provide different ways of sorting, either based on node attributes or utilizing the metrics. More details on ordering can be found in Section~\ref{sec:prototype}.}

Aggregating nodes in $N_{start}$ and/or $N_{end}$ can help to further simplify a connectivity matrix. For example, we could group nodes and their associated paths by node attributes to capture higher level phenomena in the network, to, e.g., group all airports in the New York City area. Aggregation is realized by taking the union of path sets. For instance, if two nodes $(start_{a},~start_{b})~\in~N_{start}$ are to be aggregated, then a new aggregated connectivity set is computed by taking the union of both existing sets:
\begin{equation*}
\begin{split}
C(start_{a}~\cup~start_{b},~end) = C(start_{a},~end)~\cup~C(start_{b},~end)\\
~\forall~end\in N_{end}
\end{split}
\end{equation*}
These aggregated sets can be displayed using the aforementioned metrics and encodings (Figure~\ref{fig:matrix-diagram} (right)).  Note, however, that the scales of aggregated and non-aggregated values can be quite different, which potentially requires dedicated visual encodings when showing both aggregated and not aggregated connectivity in the same matrix.

The \connectivitymatrix intentionally hides information about the intermediate nodes of paths to support analysis on the general connectivity between start and end nodes. However, understanding the role of intermediate nodes can be important for certain analysis tasks, such as identifying major hubs in a flight network. Thus, we introduce an additional, complementary visualization that focuses on the intermediate node information, which is described next. 

\begin{figure}
\centering
\includegraphics[width=\linewidth]{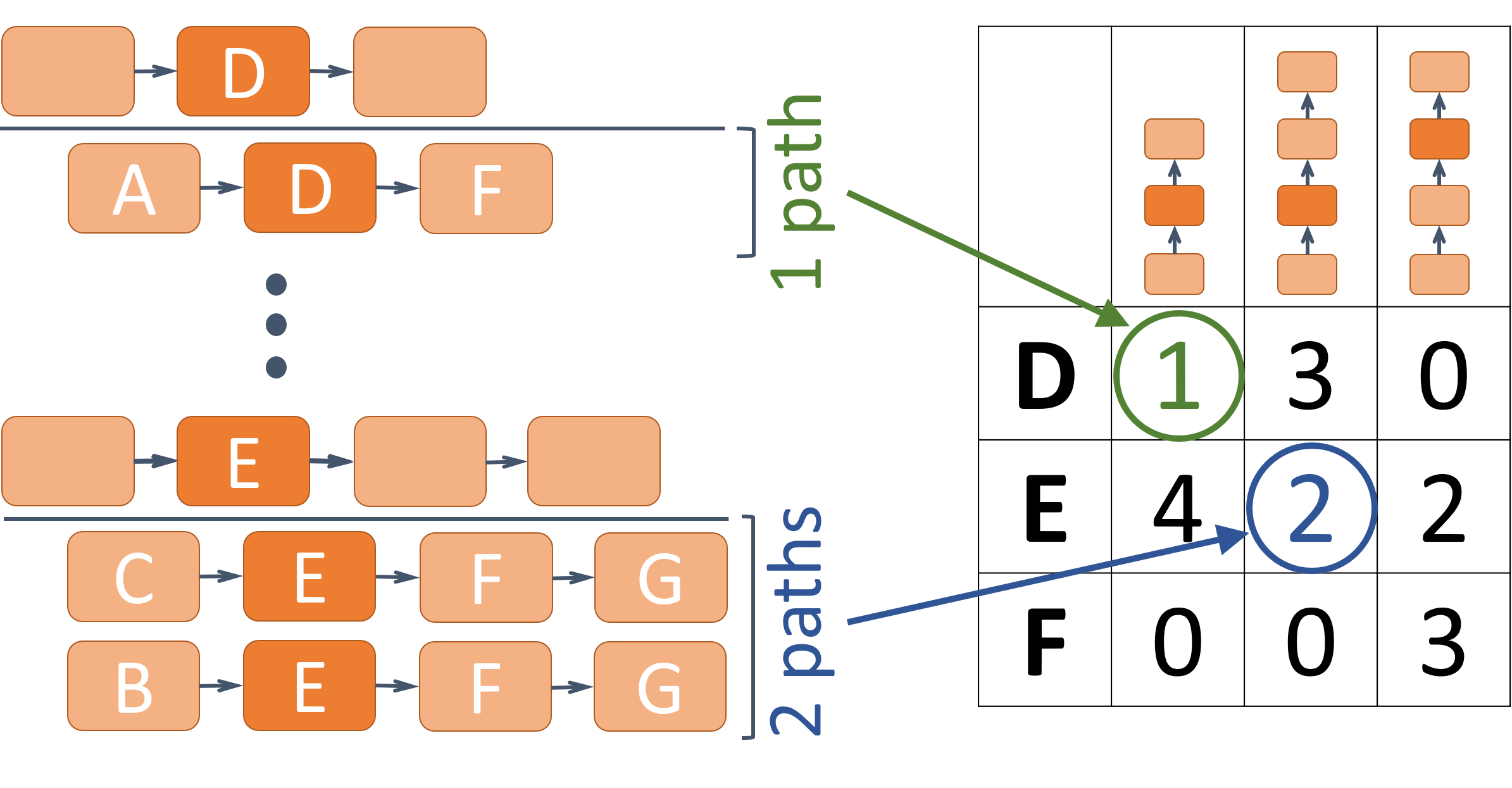}
\caption{The \intermediatenodes for the connectivity matrix shown in Figure~\ref{fig:matrix-diagram}. Rows correspond to nodes and columns correspond to a node's position in a path of a certain length. Here, node D appears once as the middle node in paths of length two. Node E is included twice as the second node in paths of length three.}
\label{fig:middle-diagram}
\vspace{-4mm}
\end{figure}

\subsection{Intermediate Node Table}

The \intermediatenodes, illustrated in Figure~\ref{fig:middle-diagram}, visualizes the properties of a path set defined by an intermediate node at a specific position in a path. \new{For instance, in the the flight graph, queries for paths of length three identify paths of three flights between four airports. The intermediate node table defines path sets based on the airports used for layovers and whether those airports are the first or second stop in the journey.} Again, we provide a formal definition of path sets and describe considerations for visualizing these sets.


Formally, the \intermediatenodes defines path sets based on the intermediate node, the path length, and node position. Let $L$ be the maximum length of all paths in the query result $P$. Let $(j, l)$ represent position $j$ in paths of length $l$. Also, let $node(p,j)$ return the node at position $j$ in path $p$. The intermediate node sets, $I$, are defined as:
\begin{equation*}
\begin{split}
I(n_{i}, (j, l))=\{p~|~p~\in~P~\wedge~Node(p, j)=n_{i}~\wedge~Len(p) =l\}\\~\forall~n_{i}~\in~N,~\forall~j~\in~\lbrack1,\dots,l\rbrack,~\forall~l~\in~\lbrack1,\dots,L\rbrack
\end{split}
\end{equation*}
These sets are represented in a table where the rows correspond to nodes and the columns correspond to the position of the node in a path of a given length. \rem{The result is that each table cell contains a metric about the path sets, e.g., it can show information about how often a node shows up in paths of different lengths.} 

\new{The number of columns in the \intermediatenodes depends on the length of paths returned by a query. In queries for paths of length two, the table contains only one column representing the middle node position in all of the paths. In queries for paths of length three, the table contains three columns representing the possible positions for nodes inside the paths as shown in Figure~\ref{fig:middle-diagram}.}

Various metrics can be used for summarizing the path sets in the \intermediatenodes. In addition to the count metric used in Figure~\ref{fig:middle-diagram}, other metrics could include the weight of paths passing through an intermediate node, or the number of unique start and end nodes that an intermediate node connects. 

Just as the \connectivitymatrix supports various visual encodings to represent metric results, the \intermediatenodes supports similar encodings. Likewise, dynamic aggregation of the intermediate node table based on node attributes is possible.


The \intermediatenodes, hence, displays a summary of the intermediate nodes returned by a path query. When paired with the \connectivitymatrix, the two techniques display an overview of paths connecting start/end nodes as well as of the importance of the intermediate nodes that those paths pass through. Interactive highlighting and selections can be used to access the relationships of paths between the two views. These interactions, along with a detailed discussion of metrics, encodings and aggregation are described in the following section.
\section{\prototypename}
\label{sec:prototype}

We have implemented the \connectivitymatrix and the \intermediatenodes in a prototype system, \prototypename (Figure~\ref{fig:full-tool}). \prototypename includes three additional components: a query interface and two supplemental views. 

While our system was designed with neuroscience data in mind, we introduce its functionality with a flight dataset. This dataset is a graph of flights in the US over three days in 2015~\cite{BTS2016}. It consists of 308 airports (nodes) and 13K flights (edges) connecting the airports. Nodes have categorical attributes, including a unique three letter airport code, a city name, and a state. \new{The categorical elements of this dataset have a hierarchical structure: one or multiple airports are associated with one city, one or multiple cities are associated with one state.} Nodes also have quantitative attributes, such as their degree, as well as geographic locations. Edges have categorical attributes, such as an identifier for the airline, and quantitative attributes, including arrival time, departure time, and length of any delays.

\subsection{Queries}

\begin{figure}[t]
 \includegraphics[width=0.7\linewidth]{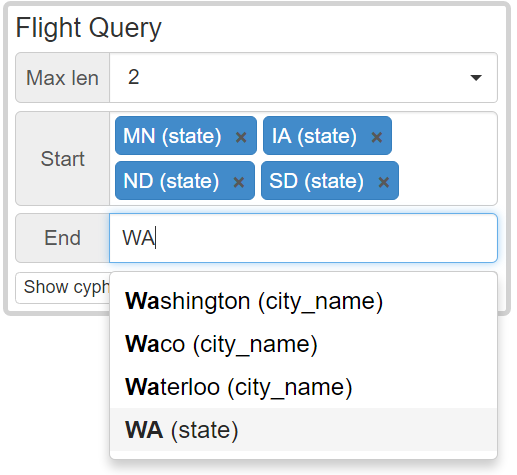}
 \centering
  \caption{The flight query interface defines paths by a maximum length as well as attributes of the start and end nodes. Here, the user can select any attributes matching the input string ``WA''.}
\label{fig:query}
\vspace{-4mm}
\end{figure}

The query interface supports visually defining queries for many-to-many paths by either specifying lists of start and end nodes or by defining node sets based on shared categorical attributes (satisfying \ref{req:query}). In addition, a maximum path length must be provided. Figure~\ref{fig:query} shows an example where the start nodes are airports in any of four states and the end nodes can be defined using any of the options shown.

\prototypename also supports defining advanced queries in the graphical interface, including restrictions on the edge types and on intermediate nodes. Examples of advanced queries are shown in the supplemental material. Additionally, queries can be specified in the cypher language~\cite{Neo2016}, which enables queries of arbitrary complexity that are not easily specified using a graphical user interface. 

In addition to queries, paths can also be filtered by quantitative or categorical node attributes. By filtering out nodes with a high degree, for example, we can reveal connections that do not go through the major hubs of a network.

\subsection{Connectivity Overview}

The connectivity overview consists of the \connectivitymatrix and \intermediatenodes as described in Section~\ref{sec:connectivity-overviews}. Here, we describe the details of their implementation, including the display of path metrics and visual encodings, dynamic aggregation, node attributes, reordering, highlighting, and selection. 

\begin{figure}
 \includegraphics[width=\linewidth]{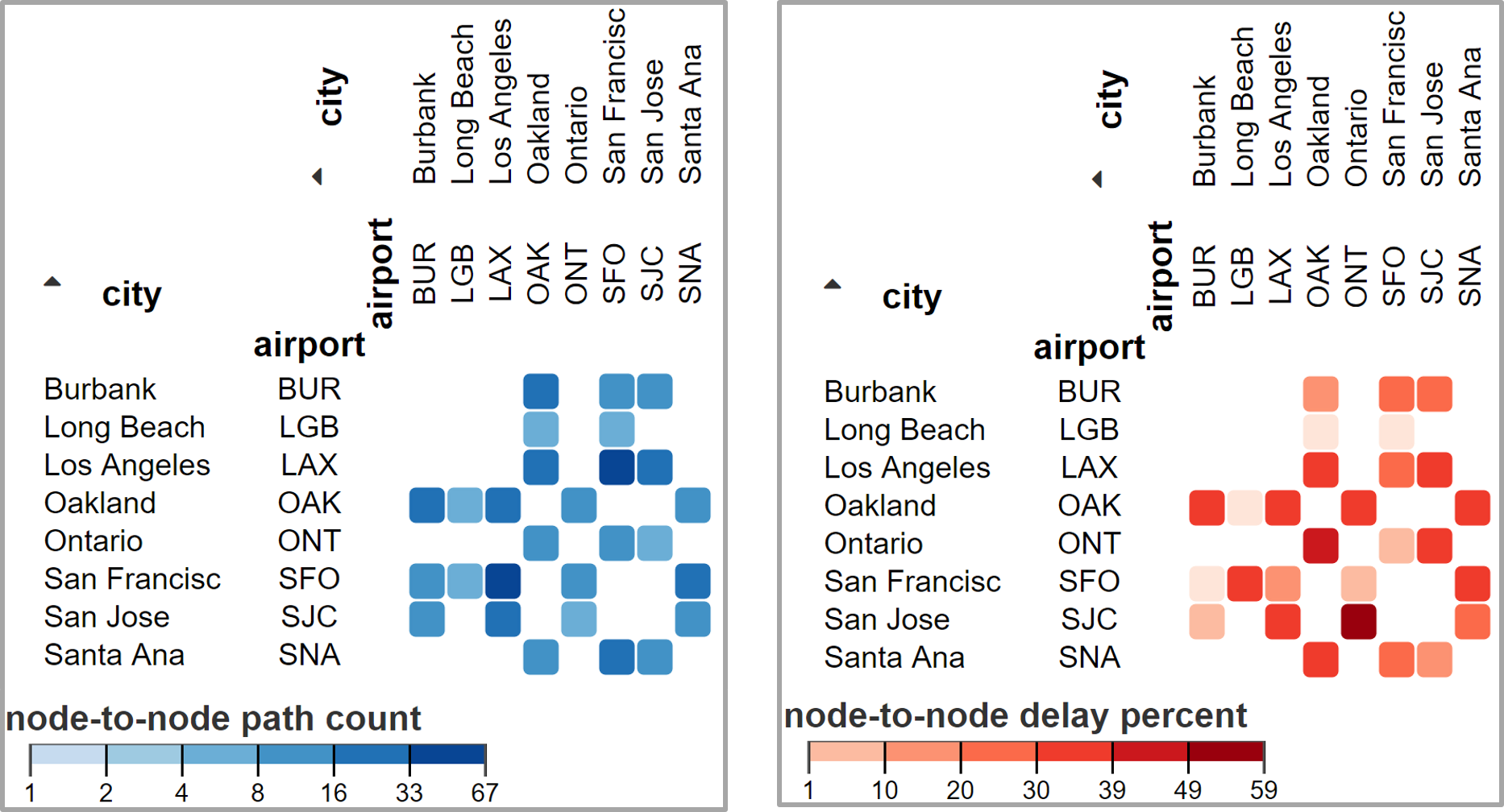}
 \centering
  \caption{Different metrics applied to direct flights between Los Angeles and San Francisco area airports. The count of flights is shown on the left, the percent of flights with more than a 15 minute delay is shown on the right.}
\label{fig:metrics}
\end{figure}

\begin{figure}
 \includegraphics[width=\linewidth]{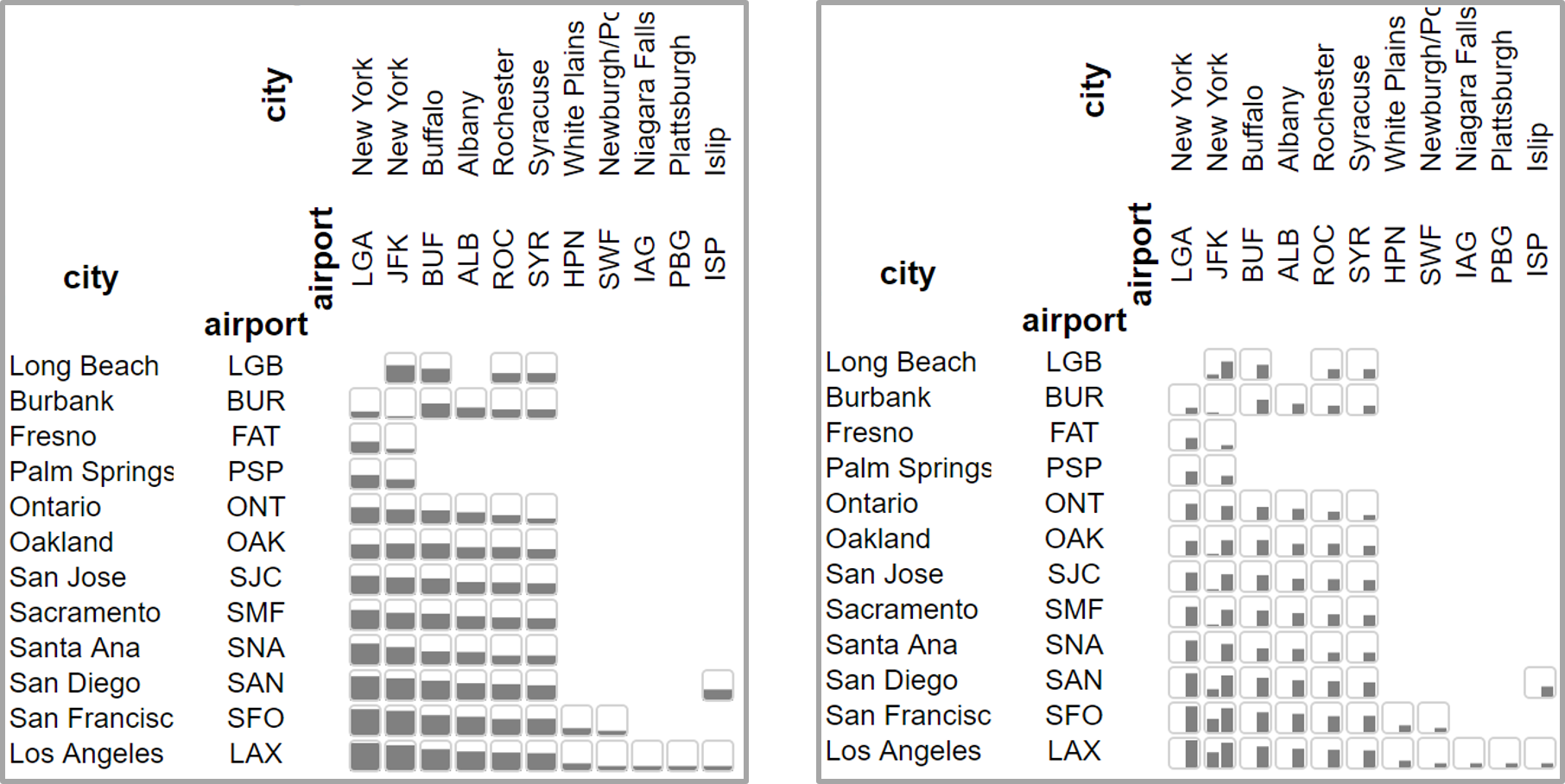}
 \centering
  \caption{Two encodings for the number of paths connecting California to New York. Left is a bar chart where height encodes the number of paths. Right is a bar chart where the left bar encodes paths of length one and the right bar encodes paths of length two.}
\label{fig:encodings}
\vspace{-5mm}
\end{figure}

\begin{figure}
 \includegraphics[width=0.9\linewidth]{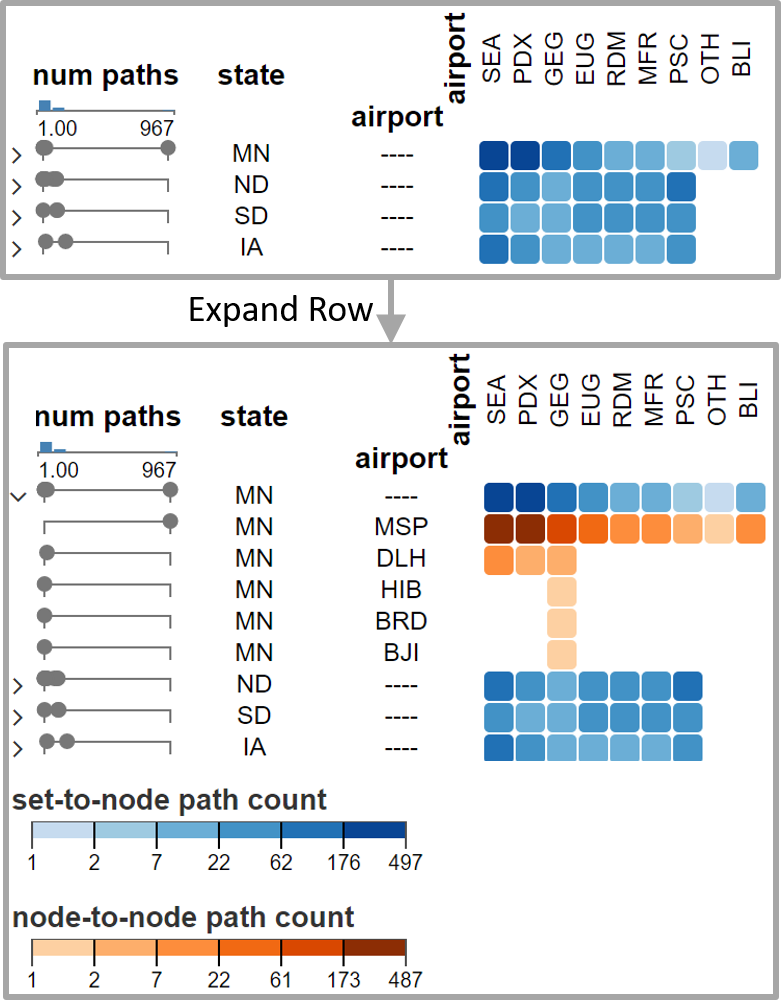}
 \centering
  \caption{The \connectivitymatrix supports dynamic aggregation of nodes by attribute. Here, the connectivity matrix from Figure~\ref{fig:full-tool} is aggregated by starting node state; the airports in Minnesota (MN) are then expanded. Different color scales in aggregated cells account for differences in scales and emphasize the aggregation. Dotplots represent quantitative attributes for both the aggregated and expanded rows.}
\label{fig:aggregation}
\vspace{-5mm}
\end{figure}

The cells in the \connectivitymatrix and the \intermediatenodes display the result of metrics applied to path sets. The default metric for both views is a path count displayed with a quantitative color map (Figure~\ref{fig:full-tool}). There are many other possible metrics beyond a path count, such as the percent of delayed flights connecting two airports, which is shown in Figure~\ref{fig:metrics}. 

In addition to dynamic metrics, Graffinity supports interactively changing the visual encodings. Figure~\ref{fig:encodings}, for example, shows two encodings that use bar charts. The left example uses a bar to encode the total number of paths. The right example contains two bars, where the first bar visualizes the number of paths of length one, and the second bar visualizes the number of paths of length two.

\prototypename supports dynamic aggregation of nodes based on their attributes. This is important for analyzing higher-level relationships in the graph, for instance, to understand connections between states instead of individual airports. This aggregation is demonstrated in Figure~\ref{fig:aggregation}, where the starting nodes are aggregated by state. Aggregated sets can be expanded to show the nested rows or columns. We use different color scales for aggregated values to (a) make it obvious that a row or column is aggregated and (b) to account for the often significantly different data ranges between aggregates and individual nodes.

\prototypename also displays node attributes. Node attributes are visualized adjacent to the rows and columns of the \connectivitymatrix and \intermediatenodes. Categorical attributes are visualized as strings. Quantitative attributes are shown using dotplots (see Figure~\ref{fig:aggregation}). \rem{The dotplots are well suited to display multiple entries, which is particularly important for representing aggregated sets of nodes. They thus enable comparisons of attributes between nodes in the same set and across sets.} \new{The dotplots are well suited to display multiple entries. This is particularly important for representing aggregated sets of nodes, such as when airports are aggregated by their state.}

\begin{figure}
 \includegraphics[width=\linewidth]{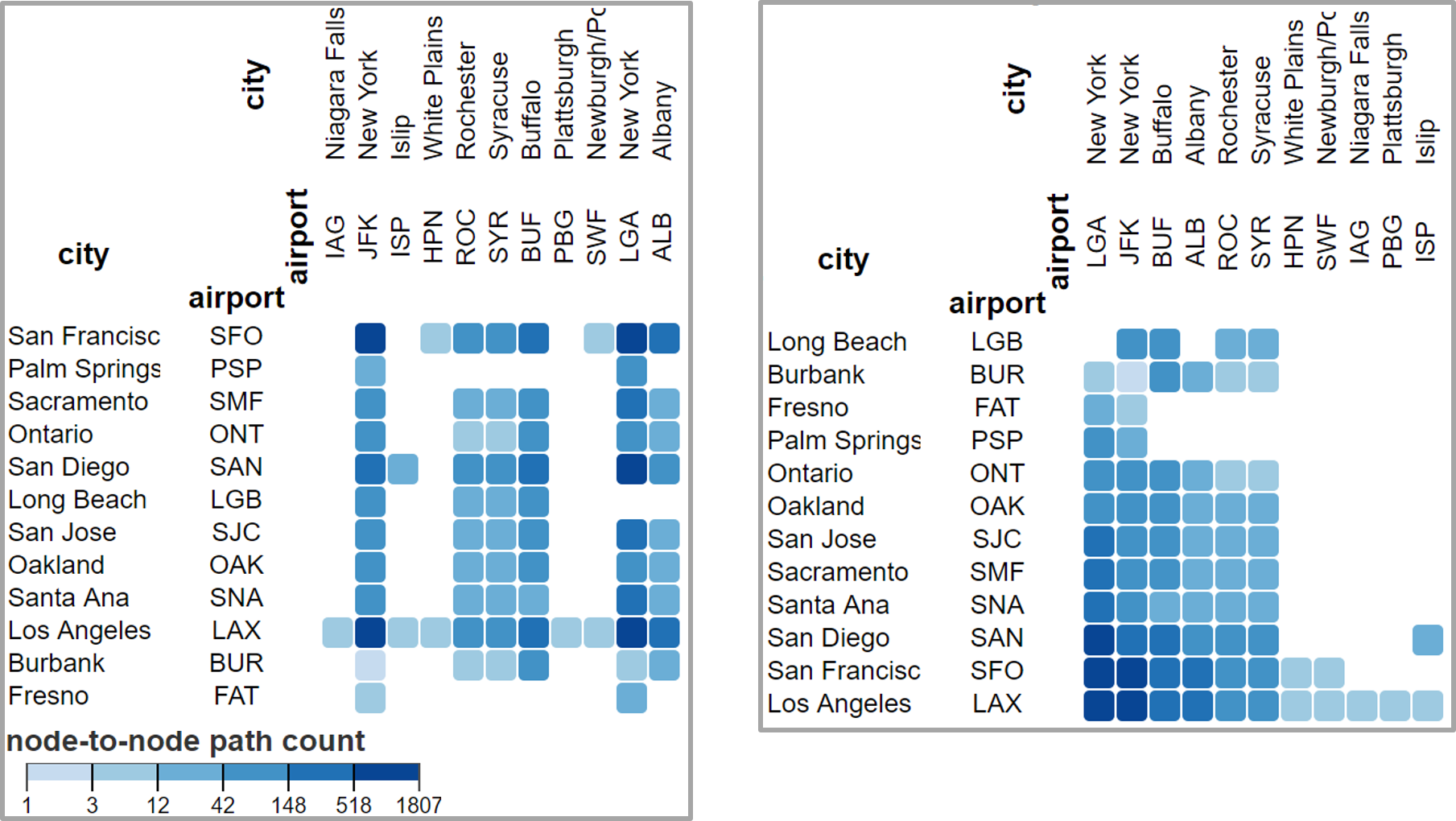}
 \centering
  \caption{The \connectivitymatrix supports dynamic ordering. The left matrix is in the order that was returned by the database query, while the right matrix is using an optimal leaf ordering algorithm.}
\label{fig:ordering}
\vspace{-5mm}
\end{figure}

The features that can be discovered in matrices are strongly influenced by the matrix ordering~\cite{Behrisch2016a}. Consequently, \prototypename supports dynamic re-ordering either based on node attributes, or using matrix reordering algorithms~\cite{Fekete2015}. An example of the optimal leaf ordering applied to a matrix is shown in Figure~\ref{fig:ordering}.

Linked highlighting reveals relationships between the \connectivitymatrix and \intermediatenodes. For example, hovering over a node or path set in the \intermediatenodes reveals the flights and paths that pass through that node in the \connectivitymatrix. Similarly, hovering over a node or path set in the \connectivitymatrix highlights the intermediate nodes used in those paths.
Individual cells can also be selected so that the contained paths can be inspected in detail in the supplemental views.

\subsection{Supplemental Views}

\begin{figure}
 \includegraphics[width=0.8\linewidth]{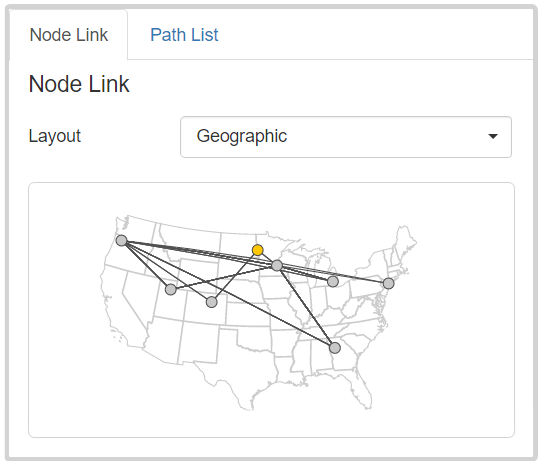}
 \centering
  \caption{A node-link view with geographic layout for the data selected in Figure~\ref{fig:full-tool}. \prototypename also supports force-directed layouts for this diagram.}
\label{fig:node-link}
\vspace{-5mm}
\end{figure}

The supplemental views are meant to provide context (\ref{req:context}) and details (\ref{req:details}) about a selection of paths. They are updated every time a cell in the \connectivitymatrix or the  \intermediatenodes is selected. \rem{We currently provide node-link diagrams and path-list views.} \new{We currently provide node-link diagrams (Figure~\ref{fig:node-link}) and path-list views (Figure~\ref{fig:full-tool}).}

We provide two layouts for the node link diagram. The first is a force-directed layout that provides topological context. It, for example, lets analyst identify well-connected nodes in the selected paths. The second layout renders the network in a spatial context and can be overlaid with, e.g., a map, as shown in Figure~\ref{fig:node-link}.

The path-list views enables analysts to browse the paths and provides details about the individual paths (\ref{req:details}). In particular, it displays a list of the selected paths in a motif hierarchy. For the flight dataset, the motifs describe the airports that flights pass through. The motifs can be expanded to display the underlying paths, e.g., to display information such as their ID, carrier, and departure times. 

The spatial layout and the motifs are domain specific, i.e., a map of the US is an appropriate layout for the US flight data, where as map of the location of neurons in a microscopy image could be an appropriate layout for the connectomics data. Similarly, the motifs and details displayed in the path list view depend on the dataset. \rem{The next section demonstrates the motifs and details used in analyzing neuroscience data.} \new{In the flight data, the airport codes provide a meaningful path aggregation while the classification of neurons provides a meaningful aggregation for our collaborators.}

\subsection{Implementation}

\prototypename is a web-based client-server tool that was developed using a combination of web technologies. The visualizations are implemented in ES6 using D3, AngularJS, and Bootstrap. The server is implemented using Python, Flask, and uses the Neo4j graph database. The path queries are executed with a breath first search strategy. We have included the source code in our supplemental material and made it available on GitHub  under an open-source license: \url{http://www.github.com/visdesignlab/graffinity}.

\section{Case Study: Retinal Connectomics}
\label{sec:validation}

\begin{figure*}
\includegraphics[width=\textwidth]{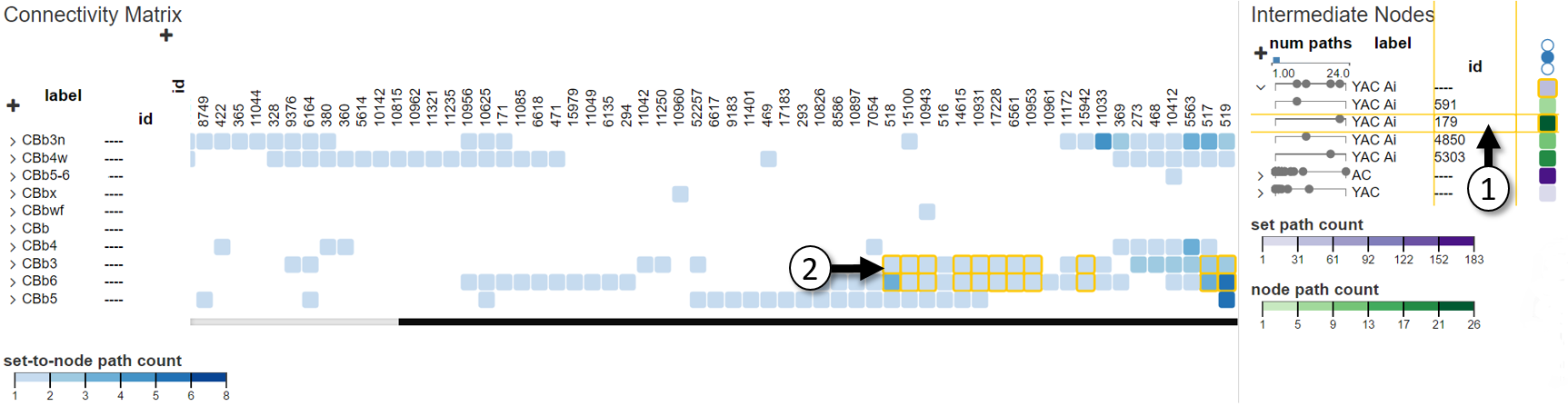}
\centering
\caption{Using \prototypename to discover anomalies in the connectome graph. Here, the connectivity matrix shows paths of length two connecting cone bipolar cells to rod bipolar cells. (1) The intermediate node 179 with label \textit{YAC Ai} participates in a large number of these crossover paths. Hovering on this row in the intermediate node table reveals the starting and ending nodes of these paths in the connectivity matrix. (2) The yellow boxes around matrix cells for the rows of \textit{CBb3} and \textit{CBb6} show that node 179 receives input from both of these classes. This is surprising as nodes with label \textit{YAC Ai} should not form connections with \textit{CBb3} nodes, though they technically could access them. We questioned whether this anomaly was a biological wiring error or a data collection error. Ultimately, Graffinity guided access to the database images, showing the anomaly to be an annotation error.}
\label{fig:crossover}
\end{figure*}


We demonstrate the usefulness of \prototypename through a case study of analyzing a \textit{connectome}, a graph of connections between cells. Our collaborators (some of whom are also co-authors) are connectomics researchers studying the connectome of cells in the retina. In this 18-month collaboration, we have leveraged user-centered design methods, such as creativity workshops~\cite{Goodwin2013} and contextual inquiry~\cite{Holtzblatt1993}, to understand the analysis needs of this group of neuroscientists. We developed \prototypename to support those needs. In this section, we briefly describe the data involved in retinal connectomics research, followed by a case study where \prototypename was used to detect errors in the connectomics dataset.

The retinal connectome that we worked with, a database called RC1, was generated from a rabbit retina through automated electron microscopy imaging, image processing, and manual annotations~\cite{Anderson2010}. It is a multivariate graph of 15K neurons (nodes) and 26K synapses (directed edges)~\cite{Anderson2011}. The nodes have categorical attributes, such as a label which specifies the type of the cell. They also have quantitative attributes, such as the size of the cell's convex hull. The edges have categorical attributes, such as the type of synapse. It is important to note that the nodes and edges in the graph are annotated based on microscopy images of the retina, i.e., the graph's nodes and edges are an abstraction of the connections observed in the images.

Understanding the connectivity of retinal cells enables researchers to reason about the flow of information through the retina and the functions of various cells. For example, Lauritzen et al.~\cite{Lauritzen2016} recently identified the winner-take-all, rod-cone crossover networks that switch between pathways for cone-driven bright light vision and those for rod-driven dim light vision. Fast crossover networks are particularly important in mesopic environments where both rods and cones are active and compete for network dominance. This circuitry was discovered through the analysis of approximately 8000 different paths of various lengths in the RC1 connectome.

In one of our sessions for getting feedback on~\prototypename, we worked with our collaborator to revisit the cone-rod crossover analysis performed by Lauritzen et al.~\cite{Lauritzen2016}. One particularly interesting part of this analysis occurred when we discovered an anomalous pathway in the dataset that had not previously been detected. In the remainder of this section, we describe the steps of detecting that anomaly and analyzing its significance --- please see our supplemental material for a more detailed set of images.

In the analysis, we queried for two-hop paths that matched the cone-rod crossover circuitry. This resulted in 272 paths that connected 90 cone bipolar cells (denoted with labels that start with \textit{CBb}) to 74 rod bipolar cells (label of \textit{Rod BC}) through 104 intermediate amacrine cells (label containing \textit{YAC} or \textit{AC}). 

In these query results, we were interested in connections formed by classes of cells. We aggregated the source nodes (rows of the \connectivitymatrix) and the intermediate nodes (rows of the \intermediatenodes) by label. We then inspected the intermediate nodes that connect rods and cones. 

In particular, we examined intermediate nodes with the label \textit{YAC Ai}. One of these cells had many more connections than the others of the same label. We expanded the aggregated \textit{YAC Ai} row and we were able to use linked highlighting between the \connectivitymatrix and \intermediatenodes to reveal the paths connected by the intermediate nodes. In particular, we noticed that cell \textit{179} received input from a cell with label \textit{CBb3} (Figure~\ref{fig:crossover}), which violated the expected connections for that cell type.

The question triggered by this finding is central to all of connectomics: is this anomaly a biological error, which addresses the nature of biological wiring precision, or a technical error inherent in connectomics mapping? By selecting the paths through cell \textit{179} in the \intermediatenodes, we were able to use the path list view to drill down to the individual synapses responsible for these paths. With these synapse IDs, we accessed the images of the database and discovered that the connection from \textit{CBb3} to \textit{YAC Ai 179} was an error. Although this crossover network had been rigorously analyzed with coarser granularity, fine-scale annotation errors persisted and these became apparent when viewed with \prototypename.


We have described one case study where \prototypename and our connectivity overview were used to support analysis in connectomics research. Our supplemental material describes an additional case study in this domain. We demonstrate how \prototypename supports analysis of communication between cell types. This provides important information for analysts who are trying to appropriately label cone bipolar cells in their database. Qualitative feedback on the prototype is included in the next section.
\section{Discussion}
\label{sec:discussion}


\rem{We designed the connectivity matrix and intermediate node table to support reasoning about connectivity relationships in large graphs.} \new{In addition to our case study validation, we discuss the qualitative feedback on Graffinity and the scalability of the proposed visualization techniques. We also reflect on the role of these techniques in the larger scope of graph analysis.}


Our collaborators provided positive feedback on the range of connectivity analysis supported by \prototypename during six hours of informal interviews and demonstrations. One analyst said that \prototypename ``generated figures that I didn't think were possible'' and that those figures were ``exactly what I need'' for her on-going research of neuron connectivity. Another analyst referred to the connectivity matrix as ``very powerful ... and truly exciting [for connectivity analysis].'' Throughout these feedback sessions we encouraged analysts to use \prototypename to visualize both novel and previously documented patterns in connectivity. In both cases, analysts were able to generate new insights about neuron connectivity.

One goal of our feedback sessions was to evaluate whether relatively short paths were sufficient for connectivity analysis. \rem{In fact, none of our collaborators were interested in analyzing paths with length greater than four.} \new{Throughout these sessions, our collaborators expressed interest in querying for paths of length four or less.} This supports our assumption that, in practice, relatively short paths are desirable for connectivity analysis, which \newrem{holds}{is certainly true} for transportation networks, and we believe is valid for many other analysis scenarios. \rem{Although the connectivity matrix scales to paths of arbitrary length, the intermediate node table works best with paths of lengths four or less. While the current visualization technique fulfills the needs of neuroscientists performing connectomics analysis, a representation of intermediate nodes that scales more effectively with path length is an interesting area for future work.} 

As the number of paths connecting two nodes increases exponentially with path length, there are computational limitations regarding query-based analysis. Path queries on the highly connected flight dataset that include paths of length three often require minutes to execute. In contrast, the neuroscience dataset is relatively sparse and supports interactive query results for paths of length four.\xspace \prototypename could be improved with streaming query results and progressive visualization updates~\cite{Fisher2012} or with heuristics that predict connectivity.

Informal testing with the flight and neuroscience datasets revealed that the \connectivitymatrix and \intermediatenodes scale effectively with the number of paths returned by a query \new{but they suffer from limitations common to other table-based visualizations as the number of nodes returned by a query increases.} Both techniques can interactively display around 100K paths as both visualizations are created in linear time. \new{However, the number of rows and columns in each visualization are limited by screen space, which can require scrolling as seen in Figures~\ref{fig:full-tool} and~\ref{fig:crossover}.}

Our reliance on queries to provide an overview of connectivity of large graphs requires that the analyst has knowledge of the graph and can formulate relevant queries. While this is true in many scenarios, such as the flight dataset, for which it is easy to formulate queries by a typical user, and for the neuroscience dataset, which our collaborators know well, it implies that \prototypename is not well suited to explore a graph that a user does not know much about. \new{Due to this, \prototypename should be used as part of a larger tool chain that supports open exploration, such as through degree-of-interest functions~\cite{VanHam2009a} or visual summaries~\cite{Wattenberg2006}.}
\section{Conclusion and Future Work}

In this paper, we introduced \new{the connectivity matrix and the intermediate node table}, two novel visualization techniques for summarizing connectivity relationships in large graphs. \new{The connectivity matrix uses the metaphor of an adjacency matrix generalized to show path-based relationships between start and end nodes. This scalable representation avoids required manual tracing of adjacency matrices. The intermediate node table reveals information about nodes hidden by the connectivity matrix. These two techniques provide an overview of tens of thousands of paths potentially using a variety of connectivity metrics.

We realized these techniques in a prototype system, called \prototypename. This system also contains two supplemental views, a path-list view and node-link diagram view, so that a wide range of connectivity questions can be answered and all our requirements can be addressed.}\rem{We implemented these techniques in a query-based system that supports dynamically querying paths, visualizes a summary of those paths, and provides path details-on-demand.} We demonstrated \prototypename's fitness for use in case studies on a retinal connectomics dataset\new{, though more work remains to integrate it into a larger tool chain for graph analysis.}

\new{Our prototype implementation illuminated interesting areas of future work focused on the exploration of connectivity metrics and visual encodings to represent these metrics.} \rem{In our prototype implementation, we introduced several connectivity metrics and visual encodings to represent these metrics.} We have demonstrated a few interesting metrics for path analysis, such as the path count and minimum length, as well as domain-specific metrics such as the percent of delayed flights. We hope to explore the design space of connectivity metrics and optimal visual encodings for their results in the future.


Finally, the \prototypename system could be extended to support comparison tasks. For example, it would be interesting to compare the flight connectivity using individual airlines, to, e.g., see the differences in connectivity of \rem{the American Airlines network and the Delta network} \new{two airline carriers}. Another interesting comparison use case is analyzing inhibitory and excitatory synaptic pathways in the retina. These comparisons could be achieved either using small multiples of the \connectivitymatrix and the \intermediatenodes, or by using explicit metrics for the differences of these queries, paired with tailored visual encodings.

\section*{Acknowledgements}
We thank the members of the Vis Design Lab and MarcLab at the University of Utah for their feedback and contributions to this work. James R. Anderson provided invaluable assistance with the RC1 connectome. This work was supported in part by NSF grant IIS-1350896, NIH U01 CA198935, the DoD - Office of Economic Adjustment (OEA) ST1605-16-01, NIH EY015128, EY02576, and EY014800 Vision Core, an unrestricted grant from Research to Prevent Blindness to the Moran Eye Center.

\bibliographystyle{eg-alpha}

\bibliography{Mendeley}

~\end{document}